\documentclass[%
 aip,
rsi,%
 amsmath,amssymb,
 reprint,%
]{revtex4-1}
\usepackage{subfigure}
\usepackage{graphicx}
\usepackage{dcolumn}
\usepackage{bm}
\usepackage{xcolor}

\begin{document}

\preprint{AIP/123-QED}
\title{Band-pass superlattice magnetic tunnel junctions}
\author{Abhishek Sharma}
\author{Ashwin. A. Tulapurkar}
\author{Bhaskaran Muralidharan}
\affiliation{Department of Electrical Engineering, Indian Institute of Technology Bombay, Powai, Mumbai-400076, India}

\begin{abstract}
We propose a high-performance magnetic tunnel junction by making electronic analogs of optical phenomena such as anti-reflections and Fabry-P\`erot resonances.  The devices we propose feature anti-reflection enabled superlattice heterostructures sandwiched between the fixed and the free ferromagnets of the magnetic tunnel junction structure. Our predictions are based on the non-equilibrium Green's function spin transport formalism coupled self-consistently with the Landau-Lifshitz-Gilbert-Slonczewski equation. Owing to the physics of bandpass spin filtering in the bandpass superlattice magnetic tunnel junction device, we demonstrate an ultra-high boost in the tunnel magneto-resistance (TMR$\approx5\times10^4\%$) and nearly 92\% suppression of spin transfer torque switching bias in comparison to a traditional trilayer magnetic tunnel junction device. The proof of concepts presented here can lead to next-generation spintronics device design harvesting the rich physics of superlattice heterostructures and exploiting spintronic analogs of optical phenomena.
\end{abstract}
\maketitle
\begin{figure}[tb!]
	\includegraphics[width=3.5in]{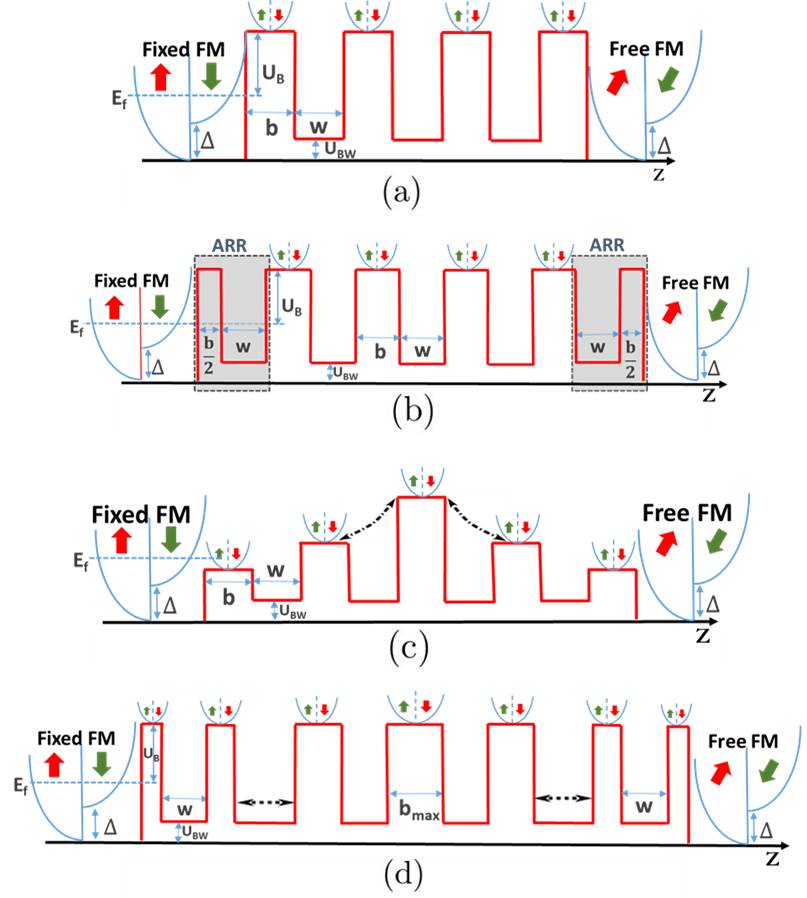}
    \caption{Equilibrium energy band profile along the $\hat{z}$ direction: (a) An SLMTJ or FPMTJ device. (b) A BP-SLMTJ-I device (also identified as AR-SLMTJ). The shaded regime is anti-reflective region (ARR) details of which has been given in supplementary material section-II. (c) Gaussian barrier height and (d) Gaussian barrier width distributed BP-SLMTJ-(II) and (III) respectively.}
	\label{band_profile}
\end{figure}
\begin{figure}[t!]
	\includegraphics[width=3.5in]{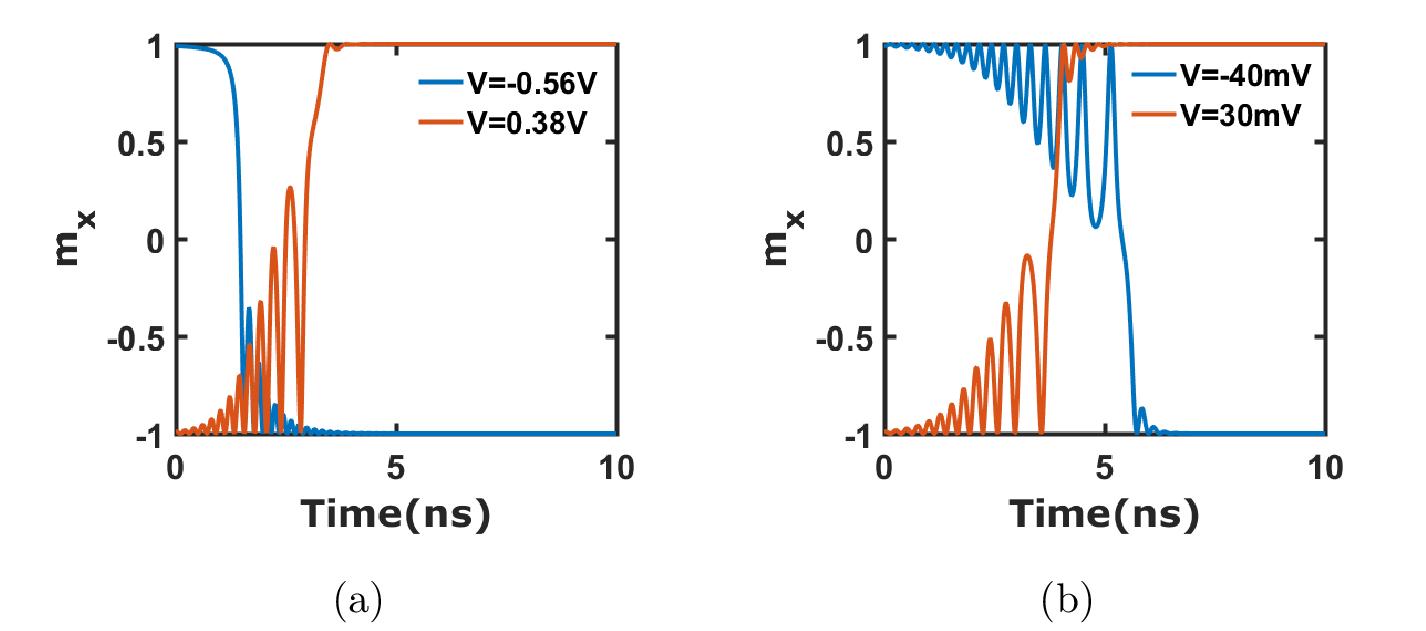}
	\caption{Spin transfer torque induced magnetization switching profiles of the free FM (a) in the trilayer MTJ device and (b) the AR-SLMTJ device with a 3-quantum well structure.}
	\label{SW_dynamics}
\end{figure}
\begin{figure}[t!]
	\includegraphics[width=3.5in]{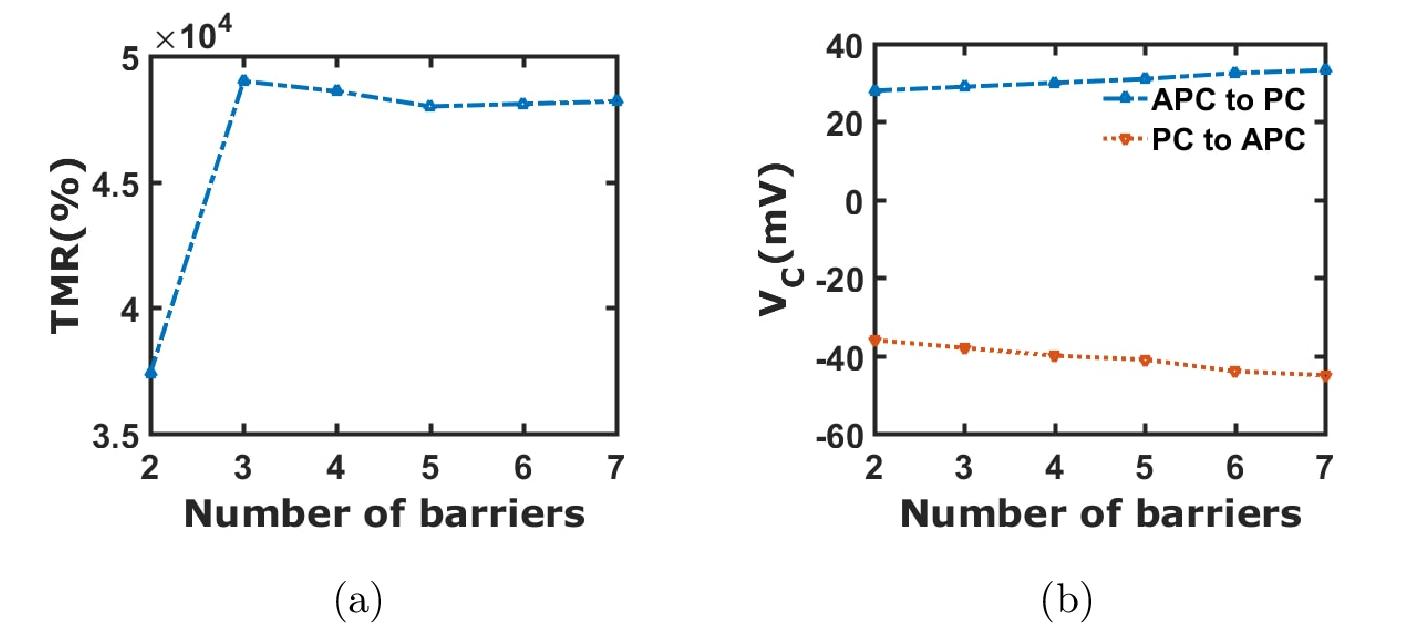}
	\caption{(a) The variation of TMR and (b) critical switching voltage ($V_C$) for the AR-SLMTJ device as a function of the number of barriers in the superlattice structure.}
	\label{TMR_SW_ncells}
\end{figure}
\begin{figure}[htb!]
	\includegraphics[width=3.5in]{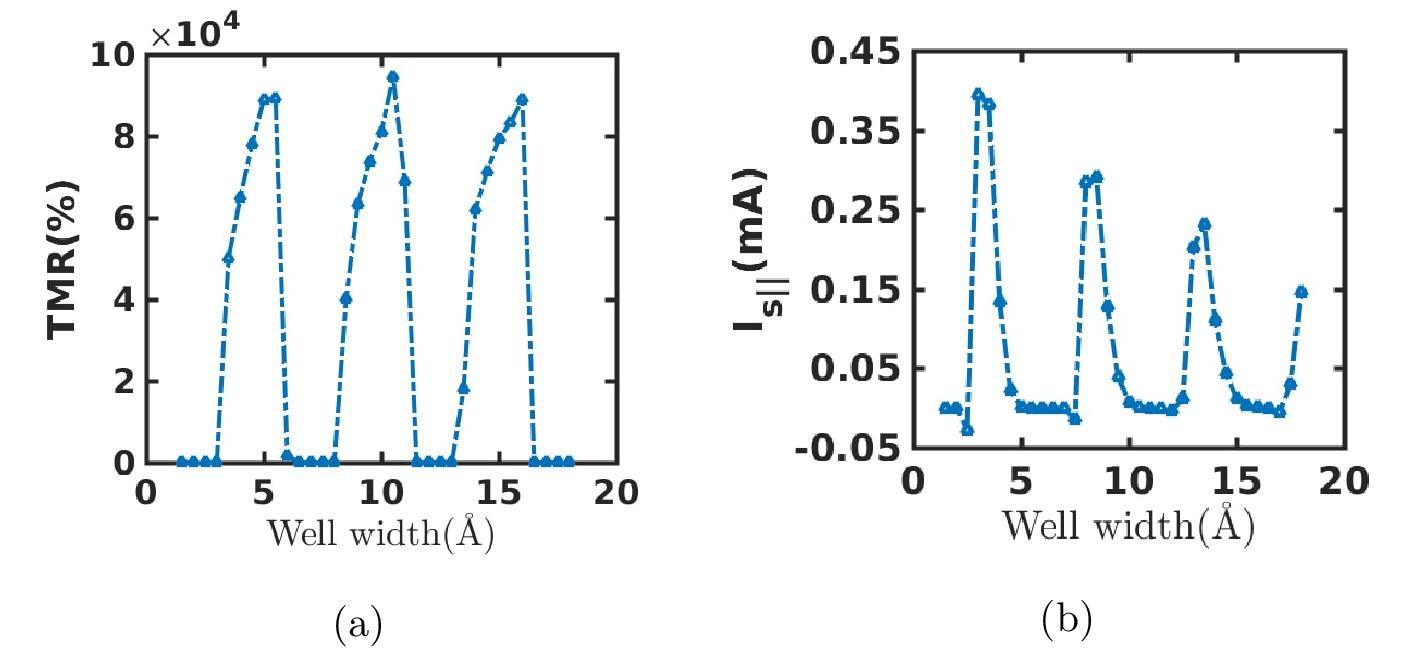}
	\caption{(a) The TMR and (b) Slonczewski's spin current as a function of quantum well width for the 3-barrier AR-SLMTJ device under an applied voltage of 20mV.}
	\label{TMR_wellW}
\end{figure}

\indent Spintronics involves the manipulation of the intrinsic spin along with the charge of electrons and has emerged as an active area of research with direct engineering applications for next-generation logic and memories. A hallmark device that leads the development of the technology is the trilayer magnetic tunnel junction (MTJ), which consists of two ferromagnets (FM) separated by an insulator such as MgO\cite{Butler2001,Parkin2004}. The MTJ structure has attracted a lot of attention due to the possibility of engineering a large tunnel magneto-resistance (TMR $\approx200\%$)\cite{Djayaprawira2005} and the current driven magnetization switching via the spin-transfer torque (STT) effect \cite{Slonczewski1996,Berger1996,Assefa2007,KhaliliAmiri2011}. Trilayer MTJs find their potential applications in magnetic field sensors \cite{VanDijken2005,Sharma2016}, STT-magnetic random access memories\cite{Kultursay2013} and spin torque nano-oscillators (STNO) \cite{Kim2012,Sharma2017}. The MTJ performance for the aforesaid applications relies on large device TMR and low switching bias \cite{Sharma2016,Sharma2017,Sharma2018}. There have been consistent efforts in terms of materials development \cite{DeTeresa1999,Yang2006,Ikeda2010} and the device structure designs \cite{Useinov2011,Chen2015,Chatterji2014a} to enhance the TMR and STT in magnetic tunnel junctions. When it comes to device structures, the double barrier MTJ has been extensively explored both theoretically and experimentally to achieve better TMR and switching characteristics \cite{Chatterji2014a,RMTJnanolatter2008}. Owing to the physics of resonant tunneling, the double barrier structure has been predicted to provide a high TMR ( $\approx2500\%$) \cite{Sharma2016,Sharma2017} and nearly $44\%$ lower switching bias  \cite{Chatterji2014a} in comparison with the trilayer MTJ device. \\
\indent The Fabry-P\`erot resonances in the electronic analog are realized by superlattice (SL) structures (Fig.~\ref{band_profile}(a)) consisting of periodic stacks of two dissimilar materials with layer thicknesses of a few nanometers. SL structures have been explored extensively in the field of photonics, electronics and thermoelectronics \cite{Brennan2000,Lin2003}. In the area of spintronics, few studies \cite{Chen2014,Chen2015} have explored SL structures made of alternate layers of an insulator and normal metal (NM) sandwiched between the two FMs as a route to enhance the TMR.\\
\indent As the principal motif of this work, we propose structures that manifest spin selective band-pass transmission spectra as a possible route to achieve superior performance MTJ devices that possess large TMR as well as low switching bias. The energy band profile of possible device structures that can be identified with such a band pass transmission spectrum are sketched in the Fig.~\ref{band_profile}(b), (c) and (d) and are termed as band pass Fabry-P\`erot/superlattice magnetic tunnel junction (BP-FPMTJ or BP-SLMTJ) I, II and III respectively. The structures when  sandwiched between two ferromagnets (FMs) can be used to achieve a spin selective band-pass transmission profile \cite{Pacher2001,Gomez1999,Tung1996}. The structure BP-SLMTJ-I (also identified as the anti-reflective  Fabry-P\`erot (superlattice) magnetic tunnel junction (AR-SLMTJ)) is a regular SL structure terminated by two anti-reflective regions (ARR) and sandwiched between the fixed and free FMs\cite{Pacher2001} (Fig.~\ref{band_profile}(b)). The BP-SLMTJ-I structures can be realized either by an appropriate non-magnetic metal sandwiched between the MgO barriers or via a heterostructure of MgO and a stoichiometrically substituted MgO $(\mbox{Mg}_\text{{x}}\mbox{Zn}_\text{{1-x}}\mbox{O})$, whose bandgap and workfunction can be tuned \cite{Li2014}. The BP-SLMTJ-II (Fig.~\ref{band_profile}(c)) is SL structure having a Gaussian variation in the barrier heights\cite{Gomez1999}. Such a structure can be realized via a stoichiometrically substituted MgO $(\mbox{Mg}_\text{{x}}\mbox{Zn}_\text{{1-x}}\mbox{O})$ whose barrier height can be tuned by changing the Zn mole fraction. The well regime in the BP-SLMTJ-II structure can be realized either via a non-magnetic metal or a lattice matched ZnO\cite{Shi2012}. The BP-SLMTJ-III (Fig.~\ref{band_profile}(d)) structure is based on a Gaussian distribution of the widths of the MgO barriers in a typical SL structure\cite{Tung1996}. This can be realized either by an appropriate non-magnetic metal sandwiched between the MgO barriers or via a heterostructure of MgO and stoichiometrically substituted MgO $(\mbox{Mg}_\text{{x}}\mbox{Zn}_\text{{1-x}}\mbox{O})$ whose band offsets can be tailored \cite{Li2014}. \\ 
\indent To establish the proof of our concept, we present here a detailed analysis of BP-SLMTJ-I or AR-SLMTJ that incorporate electronic analogs of optical phenomena such as anti-reflection coatings (ARC) and Fabry-P\`erot resonances. We demonstrate that owing to the bandpass spin-filtering physics of the BP-SLMTJ structure, the proposed AR-SLMTJ device exhibits large non-trivial spin current profiles along with an ultra-high tunnel magnetoresistance, leading to an enhanced switching performance.\\
\begin{figure}[t!]
	\includegraphics[width=3.5in]{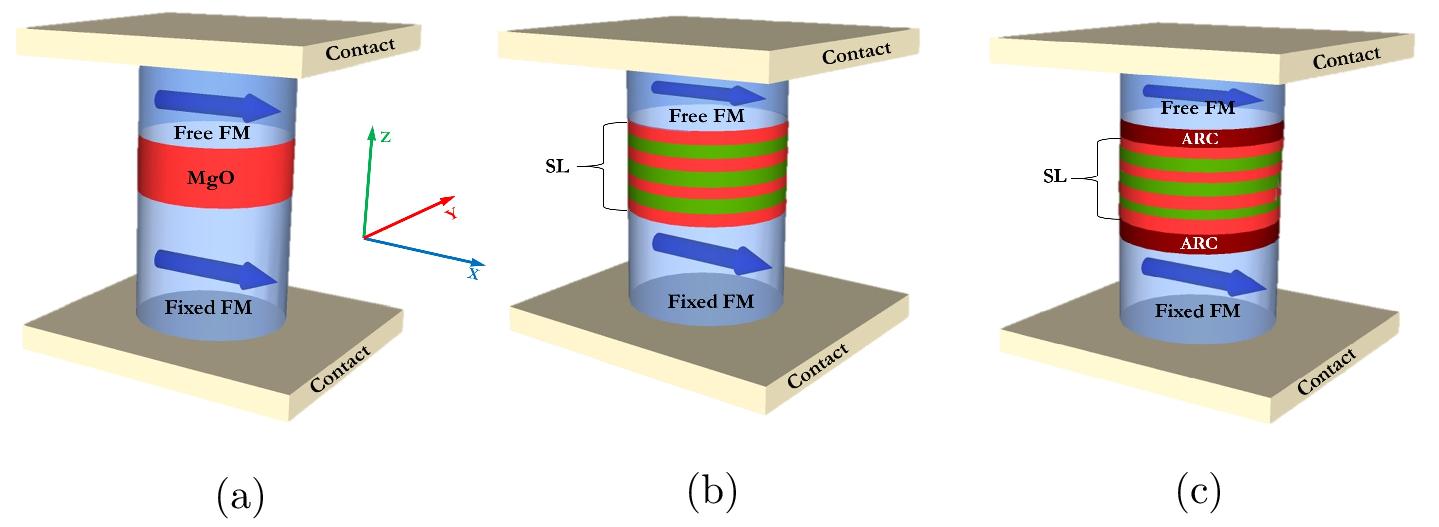}
	\caption{Device schematics: (a) A trilayer magnetic tunnel junction (MTJ) device having a MgO barrier separating fixed and free FM layers, (b) a  SLMTJ with 4-barriers or 3-quantum wells having alternating layers of MgO (red) barrier and normal metal (green) well sandwiched between the free and the fixed FM layers, (c) the AR-SLMTJ device comprises of a superlattice heterostructure along with anti-reflection regions sandwiched between the free and the fixed FM layers.}
	\label{device_design}
\end{figure}
\begin{figure}[tb!]
	\includegraphics[width=3.5in]{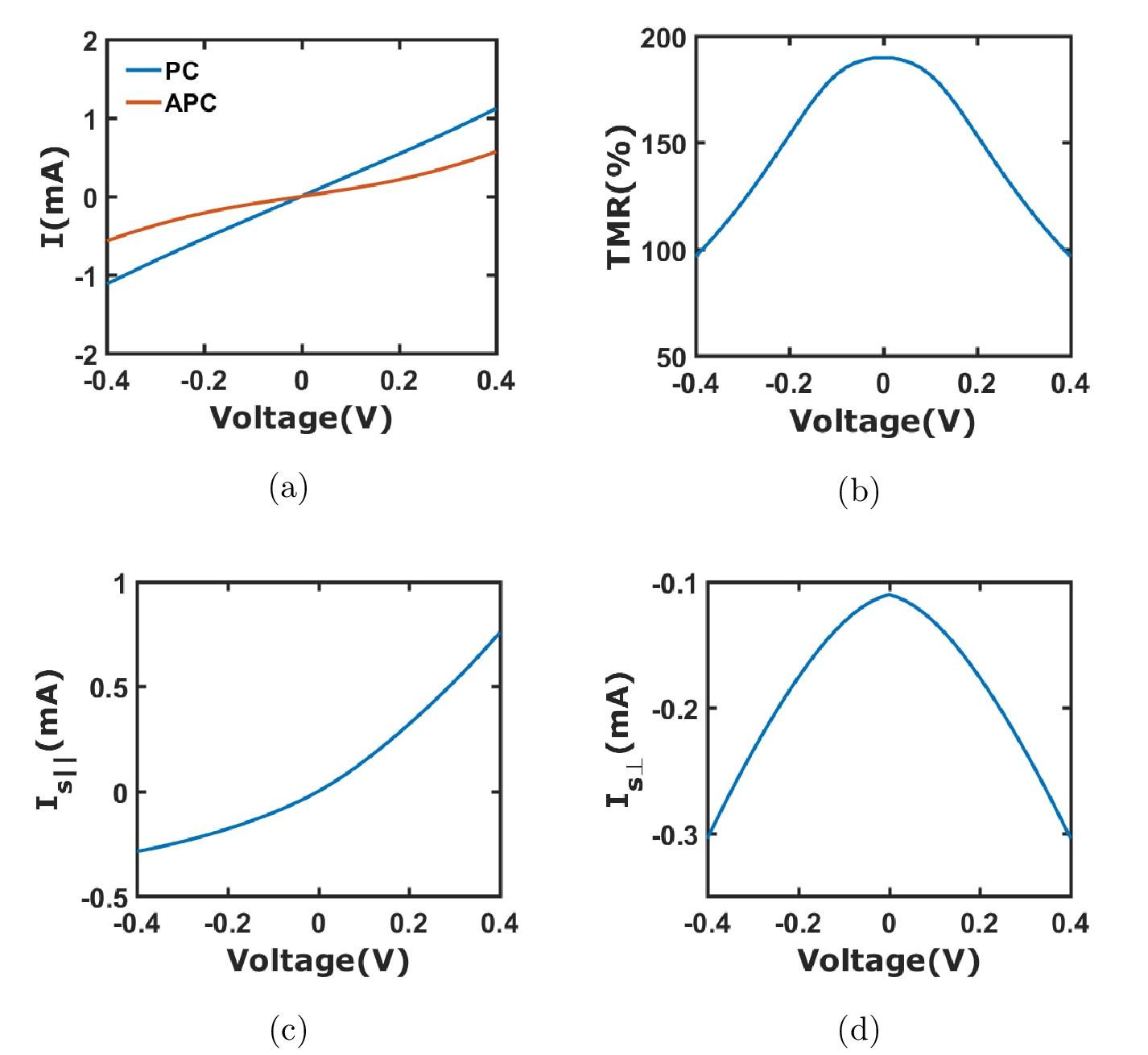}
	\caption{Trilayer MTJ device characteristics: (a) I-V characteristics in the PC and the APC, (b)  TMR variation with bias voltage, (c) variation of $I_{S\parallel}$ (Slonczewski term) and (d)  variation of $I_{S\perp}$ (field-like term) with applied voltage in the perpendicular configuration of the free and fixed FMs.}
	\label{TMR_V}
\end{figure}
\indent We show in Fig.~\ref{device_design}(a), the device schematic of a typical trilayer MTJ. Device schematics for both the SLMTJ and the AR-SLMTJ structures are depicted in Fig.~\ref{device_design}(b), and Fig.~\ref{device_design}(c), respectively. We show in Fig.~\ref{band_profile}(a), and Fig.~\ref{band_profile}(b), the band profile schematics of the SLMTJ and the AR-SLMTJ, respectively. The anti-reflective (AR) region is a quantum well and a barrier structure, whose well width is the same as that of the SL well and has a barrier width of half the SL barrier width, as depicted in Fig.~\ref{device_design}(d). The AR in a SL structure is analogous to an optical ARC that exploits the wave nature of the electrons. The electronic AR region is designed to get a perfect transmission at a particular energy, and simultaneously enhancing the transmission in the entire miniband. We have employed the non-equilibrium Green's function (NEGF) \cite{datta1} spin transport formalism coupled with the Landau-Lifshitz-Gilbert-Slonczewski (LLGS)\cite{Slonczewski1996} equation to describe magnetization dynamics of the free FM to substantiate our designs. Details of the calculations are sketched out in the supplementary material (SM) section-I.\\ 
\indent In our simulations, we use CoFeB as the FM with its Fermi energy, $E_f = 2.25$eV and exchange splitting $\Delta = 2.15$ eV. The effective mass of MgO, the normal metal (NM) and the FM are $m_{OX} = 0.18m_e$, $m_{NM} = 0.9m_e$ and $m_{FM} = 0.8m_e$, respectively \cite{deepanjan}, with $m_e$ being the free electron mass. The barrier height of the CoFeB-MgO interface is $U_B = 0.76$ eV above the Fermi energy \cite{deepanjan,kubota}. The conduction band offset of the NM and from the FM band edge is $U_{BW} = 0.5$ eV. We have used a barrier width of $1.2$nm chosen such that half of the barrier width is $0.6$nm which is the minimum amount of MgO that can be deposited reliably \cite{Deac2008}. The quantum well has a width of $3.5 \AA$ which is very well within the current fabrication capabilities\cite{Ryu2013,Yang2015}. It must be noted that resonant effects in metallic quantum wells are low temperature phenomena that have been observed experimentally in double barrier resonant structures with ferromagnetic contacts \cite{RMTJnanolatter2008}.\\
\indent In the results that follow, the parameters chosen for the magnetization dynamics are $\alpha$ = 0.01, the saturation magnetization, $M_S=1100$ emu/cc, $\gamma$ = 17.6 MHz/Oe, uni-axial anisotropy, $K_{u2}=2.42\times10^4$erg/cc along the $\hat{x}$-axis and the demagnetization field of $4$$\pi$$M_s$ along the $\hat{z}$-axis of the free FM\cite{deepanjan}. The cross-sectional area of all the devices considered is 70 $\times$ 160 nm\textsuperscript{2} with the thickness of the free FM layer taken to be $2$ nm. The critical spin current required to switch the free FM as described by the above parameters is around $I_{sc}\approx0.52$mA\cite{Ralph1}.\\
\begin{figure}[tb!]
	\includegraphics[width=3.5in]{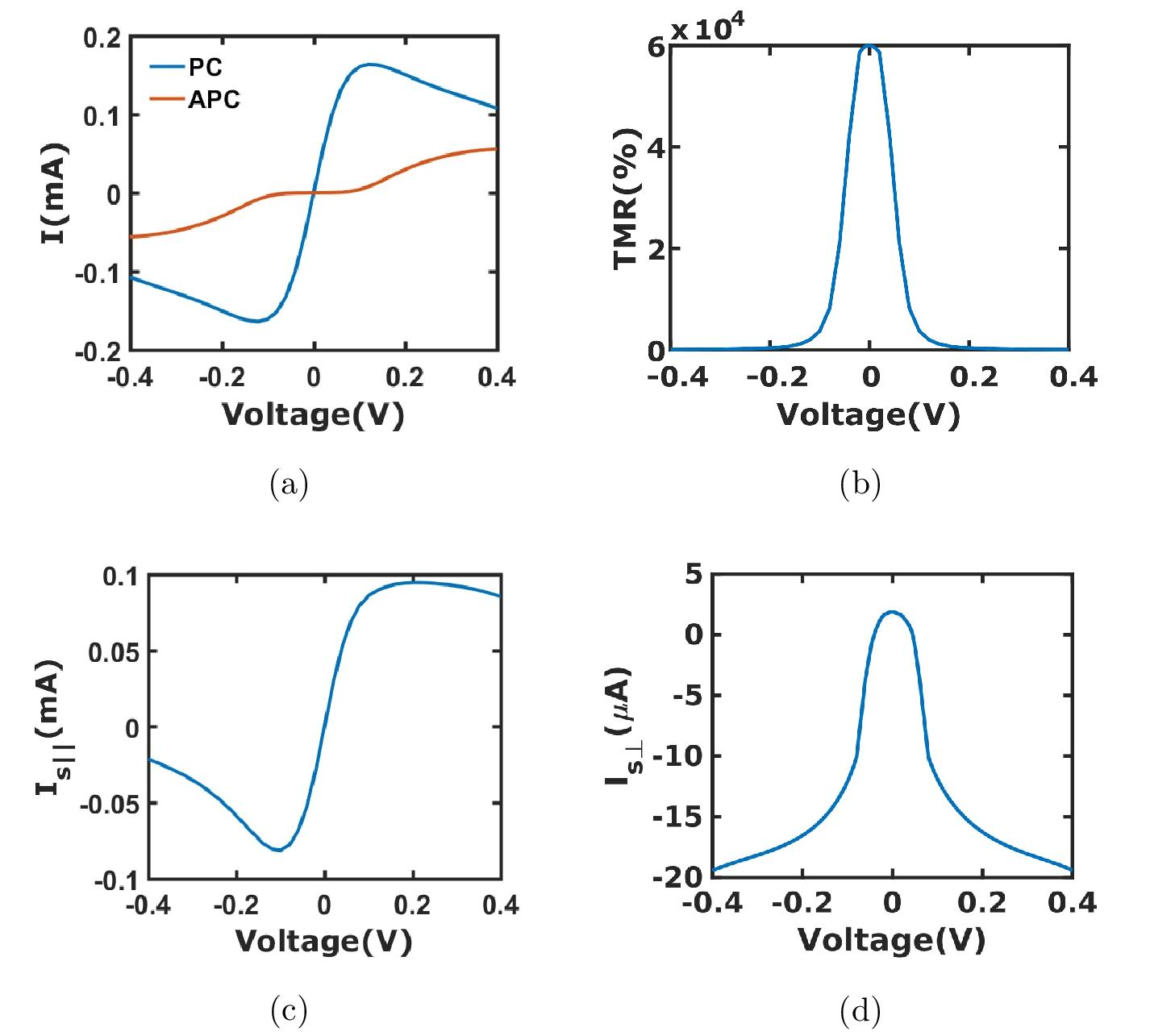}
	\caption{SLMTJ device characteristics: (a) I-V characteristics in the PC and the APC, (b)  TMR variation with applied voltage, (c) variation of $I_{S\parallel}$ (Slonczewski term) and (d)  variation of $I_{S\perp}$ (field-like term) with applied voltage in the perpendicular configuration of the free and fixed FMs.}
	\label{SLTMR_V}
\end{figure} 
\indent Spin-dependent tunneling in spintronic devices results in different amounts of charge currents flowing in the parallel configuration (PC) and the anti-parallel configuration (APC) of the FMs at a given applied bias. Figure~\ref{TMR_V}(a) shows the current-voltage (I-V) characteristics of a trilayer MTJ device in the PC and APC. Spin dependent charge flow is quantified by the tunnel magnetoresistance (TMR),  defined as $TMR=(R_{AP}-R_P)/(R_P)$, where $R_P$ and $R_{AP}$ are the resistances in the PC and the APC, respectively. The TMR variation with the voltage for a trilayer device is shown in the Fig.~\ref{TMR_V}(b). The spin current is a rate of flow of angular momentum that can act as a torque on the magnetization of the free FM. The spin current can be resolved into two components, namely, the Slonczewski term ($I_{S\parallel}$) and the field like term ($I_{S\perp}$) depending on effects of different magnitudes of the spin currents on the magnetization dynamics of the free FM.   We show in Fig.~\ref{TMR_V}(c), the variation of the Slonczewski term \cite{butler} ($I_{S\parallel}$) of the spin current with bias voltage. The Slonczewski term can either act as a damping term or as an anti-damping term in the magnetization dynamics of the free FM, regulated by the direction of the charge current. When the Slonczewski term acts as an anti-damping term in the magnetization dynamics, it can destabilize the magnetization of the free FM and can result in the switching of the free FM magnetization direction. Figure~\ref{TMR_V}(d) shows the variation of the field-like term\cite{butler} ($I_{S\perp}$) of the spin current with voltage bias. The field-like term of the spin current acts like an effective magnetic field in the magnetization dynamics and can switch the free FM. The non-vanishing part of the field-like term at zero-bias is a dissipationless spin current and represents the exchange coupling between the FMs due to the tunnel barrier \cite{Slonczewski1996}. The nature of the exchange coupling is determined by the relative positioning of the conduction bands in the FM layers and the insulator. In an MgO based trilayer device sandwiched between CoFeB FM layers, the exchange coupling is of anti-ferromagnetic nature.\\
\indent We show in Fig.~\ref{SLTMR_V}(a), the I-V characteristics of the SLMTJ with 4-barriers/3-quantum well structure in the PC and APC. The I-V characteristics depict a considerable difference between the PC and APC, which results in an ultra-high TMR as shown in the Fig.~\ref{SLTMR_V}(b). The TMR shows a roll-off with voltage bias and is attributed to the voltage dependent potential profile across the superlattice structure \cite{deepanjan}. Figure~\ref{SLTMR_V}(c) shows the variation of the Slonczewski term $I_{S\parallel}$ of the spin current with voltage bias. The Slonczewski term increases and acquires the maximum value of $I_{S\parallel}\approx0.1mA$ and then starts to fall with bias due to the off-resonance conduction. The largest value of $I_{S\parallel}\approx0.1mA$ in the SLMTJ is nearly five times smaller than the critical spin current required for magnetization switching in the free FM via the spin transfer torque (STT) effect \cite{Kim2012}. While the SLMTJ has an ultra-high TMR but smaller spin currents positions the SLMTJ as an unfavorable choice for STT switching. Although SLMTJ can be designed to provide a large spin current by having an allowed band of the transmission spectrum within energy range between $\Delta$ and $E_f$, the device design yields a very low TMR value\cite{Chen2017}. The  $I_{S\perp}$ (field-like term) variation with voltage bias is shown in Fig.~\ref{SLTMR_V}(d), and it can be inferred from Fig.~\ref{SLTMR_V}(d) that the field-like term here is negligible to induce any significant magnetization dynamics of the free FM.\\
\begin{figure}[t!]
	\includegraphics[width=3.5in]{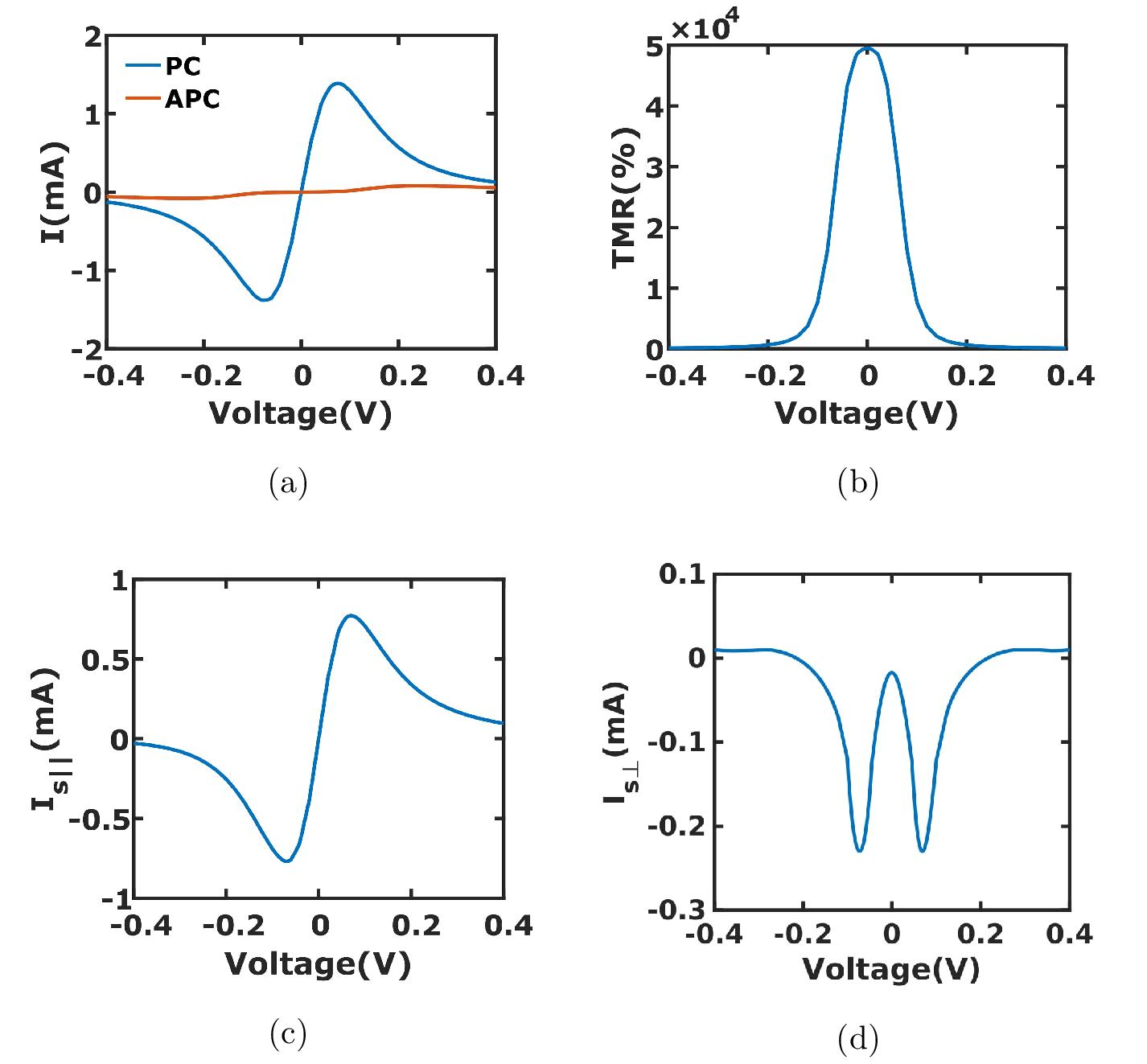}
	\caption{AR-SLMTJ device characteristics: (a) I-V characteristics in the PC and the APC, (b)  TMR variation with bias voltage, (c)  variation in $I_{S\parallel}$ (Slonczewski term) and (d)  variation in $I_{S\perp}$ (field-like term) with applied voltage in the perpendicular configuration of the free and fixed FMs.}
	\label{ARC_SLTMR_V}
\end{figure}
\indent We now plot the I-V characteristics for the AR-SLMTJ with a 4-barrier/3-quantum-well structure in Fig.~\ref{ARC_SLTMR_V}(a) in the PC and the APC. The AR-SLMTJ shows a significant asymmetry in the current conduction in both the PC and the APC which manifests as an ultra-high TMR across the structure. Figure~\ref{ARC_SLTMR_V}(b) shows the TMR variation for AR-SLMTJ with voltage bias, which is seen to have the same order of magnitude as the TMR of the SLMTJ near zero bias. An ultra-high TMR in the SLMTJ and AR-SLMTJ is ascribed to physics of spin selective filtering described in the SM section-IV. We show in the Fig.~\ref{ARC_SLTMR_V}(c), the variation of the Slonczewski term $I_{S\parallel}$ of the spin current with the voltage bias. The Slonczewski term $I_{S\parallel}$ in the AR-SLMTJ shows a nearly symmetric behavior around zero bias which may enable a near symmetric switching bias in this device. It can be seen clearly from Fig.~\ref{ARC_SLTMR_V}(c), \ref{SLTMR_V}(c) and \ref{TMR_V}(c) that the AR-SLMTJ provides a large spin current in comparison to the SLMTJ and the trilayer MTJ due to the physics of selective band-pass spin filtering. We have also rationalized the enhance STT in the AR-SLMTJ structure via the analysis of the Slonczewski spin current transmission described in the SM section-IV. We show in the Fig.~\ref{ARC_SLTMR_V}(d), the  $I_{S\perp}$ (field-like term) variation with the voltage bias. The field-like term in the AR-SLMTJ is small and have been neglected to evaluate switching biases (see SM section-I) .\\
\indent We show in Fig.~\ref{SW_dynamics}, the temporal variation in the $\hat{x}$ component of the magnetization vector of the free FM layer due to the spin transfer torque at a voltage bias slightly higher than the critical switching voltage. It can be inferred from Fig.~\ref{SW_dynamics}(a) that APC to PC switching (red) for a trilayer MTJ device is induced by the Slonczewski term which signals an unstable oscillation in the magnetization dynamics before switching. The magnetization switching from PC to APC in a trilayer device is difficult to achieve through the Slonczewski term due to the asymmetry in negative bias and hence can be facilitated by field like terms. The magnetization switching from the PC to APC (blue) is attributed to the field-like term as shown in the Fig.~\ref{SW_dynamics}(a) due to its temporal variation during switching. The AR-SLMTJ device shows nearly symmetric variation in the Slonczewski term with the bias around zero bias. The symmetric Slonczewski term and a small field-like term in the AR-SLMTJ facilitates the APC to PC and PC to APC switching via the Slonczewski term itself as shown in Fig.~\ref{SW_dynamics}(b). A different switching voltage bias is required to switch from APC to PC and PC to APC due to the angular dependence of the Slonczewski term in the AR-SLMTJ device.\\
\indent The superlattice structure is identified by the number of alternate quantum barriers and wells. The number of peaks in the transmission spectrum of a superlattice is either equal to the number of quantum wells or one less than the number of barriers in the SL structure (see SM section-II). We show in Fig.~\ref{TMR_SW_ncells}(a), the TMR variation with the number of barriers in the superlattice of the AR-SLMTJ device. The TMR increases with an increase in the number of barriers as the transmission spectrum transitions from unity to nearly zero value with increase in the number of barriers (see SM section-II). The TMR eventually saturates with the number of barriers as the transition in its transmission spectrum approaches a step function. Figure.~\ref{TMR_SW_ncells}(b) shows that the critical switching bias increases with an increase in the number of barriers.  In the AR-SLMTJ structure, an increase in the number of barriers increases the fluctuation in the band-pass spectra of transmission which reduces the band-pass area under the transmission spectra to contribute in spin and charge flow. This increases the critical bias voltage requirement for magnetization switching due to spin transfer torque. It can be seen from the Fig.~\ref{TMR_SW_ncells}(b) that the critical switching voltage strength for APC to PC switching is lesser than that of PC to APC due to the angular dependence of the Slonczewski term in the AR-SLMTJ device. We can also infer from the above discussion that there is nearly 92\% and 92.8\% decrease in the switching bias from APC to PC and PC to APC respectively, in the AR-SLMTJ device in comparison to the traditional trilayer MTJ device.\\       
\indent We show in Fig.~\ref{TMR_wellW} the effect of quantum states of the AR-SLMTJ structure on the TMR and Slonczewski spin current. The variation in the width of the quantum wells in the AR-SLMTJ structure changes the position of the transmission spectrum with respect to the Fermi-level and manifests as a periodic variation in the TMR as a function of the well width as seen in Fig.~\ref{TMR_wellW}(a). Figure~\ref{TMR_wellW}(b) shows the variation of the Slonczewski spin current as a function of the well width. Due to the quantum states of the structure, the spin current also shows a periodic variation with the quantum well width. It can be inferred from the Fig.~\ref{TMR_wellW} that the width of the quantum well at which either the largest TMR or the highest Slonczewski current is observed does not converge to singular points. But still, in the design landscape of the well width, there are many possibilities which facilitate the AR-SLMTJ device design with a boosted TMR and low switching bias.       
\\
\indent We have proposed a fresh route for high-performance spin-transfer torque devices by tapping the band-pass transmission profile of an AR-SLMTJ structure sandwiched between two the two FM layers. We showed that the physics of spin selective band-pass filtering, enabled through the AR region translates to an ultra-high TMR with ultra-low switching bias. We have estimated that the AR-SLMTJ device caters to a TMR$ (\approx5\times10^4\%)$ and nearly to a $92\%$ lowering of the switching bias in comparison to a typical trilayer MTJ device. We believe that our idea of using band-pass transmission engineering will open up further theoretical and experimental endeavors in spintronics field. Specifically, it would be interesting to investigate the BP-SLMTJ structures to provide for enhanced thermal spin-transfer torque \cite{Bauer_TSTT} by engineering ``box-car'' spin selective transmission profiles \cite{Whitney2014}. The idea of bandpass spin-filtering can also be extended to similar device structures for ``multilevel spin transfer torque devices''\cite{multilevelSTT2018}.
\\{\it{Supplementary Material:}} See supplementary material for details about calculations, anti-reflective region design, Slonczewski spin current transmission and physics of spin filtering.
\\{\it{Acknowledgements:}} The author Abhishek Sharma would like to acknowledge Smarika Kulshrestha for her suggestions on the initial draft of this work.

\bibliography{aipsamp}

\end{document}